\documentstyle{europhys} 
\def\etal{{\hbox{{\tenit\ et al.\/}\tenrm :\ }}}

\def\And{{\rm and\ }}

\def\stars{\bigskip\centerline{***}\medskip}

\newif\ifboo \boofalse

\def\Review#1{\boofalse{\it #1},}
\def\Name#1{{\sc #1},}
\def\Vol#1{\ifboo Vol. {\bf #1}\else{\bf #1}\fi}
\def\Year#1{\ifboo #1\else(#1)\fi}
\def\Book#1{\bootrue{\it #1},}
\def\Page#1{\ifboo {\rm p. #1}\else{\rm #1}\fi}
\newcommand{\text}[1]{\hbox{\rm #1}} 
\renewcommand{\vec}[1]{\mbox{\boldmath ${#1}$}} 

\begin{document} 
\euro{}{}{}{} 
\Date{} 
\shorttitle{A. KN\"ABCHEN \etal ABSORPTION OF SAW'S BY QUANTUM DOTS} 

%\bibliographystyle{prsty} 
%\large
\title{Absorption of surface acoustic-waves by quantum dots:\\ 
discrete spectrum limit}

\author{Andreas Kn\"abchen\inst{1}, Ora Entin-Wohlman\inst{2},  
Yuri Galperin\inst{3}, \And Yehoshua Levinson\inst{1}} 
 
\institute{ 
\inst{1} 
 Weizmann Institute of Science, 
Department of Condensed Matter Physics, 
Rehovot 76100, Israel\\ 
\inst{2} School of Physics and Astronomy,  
Raymond and Beverly  
Sackler Faculty of Exact Sciences, 
Tel Aviv University, Tel Aviv 69978, Israel\\ 
\inst{3} Department of Physics, University of Oslo, P.\ O.\ Box 
1048 Blindern, 0316 Oslo, Norway} 
 
\rec{}{in final form } 
 
\pacs{ 
\Pacs{72}{50.+b}{Acoustoelectric effects} 
\Pacs{73}{20.Fz}{Weak localization effects}  
} 
 
\maketitle 
%\large
\begin{abstract}
The absorption of surface acoustic waves (SAW's) by an array of 
quantum dots in which the mean level spacing  
$\Delta$  is larger than the sound frequency  $\omega$, the temperature $T$, and the  
phase breaking rate $\tau^{-1}_\phi$ is considered. 
The direct and the intra-level (Debye) contributions to the SAW attenuation coefficient $\Gamma$ are 
evaluated, and it is shown that the sensitivity to weak magnetic 
fields and spin-orbit scattering  
(``weak localization effects'') is dramatically enhanced as compared to the case 
of a continuous spectrum, $\Delta < \tau^{-1}_\phi$. It is argued that the non-invasive measurement of $\Gamma$
represents a new tool for the investigation of the temperature dependence of the  energy relaxation rate, 
$\tau^{-1}_\epsilon$, and the phase breaking rate, $\tau^{-1}_\phi$, of isolated electronic systems. 
\end{abstract} 
% 72.50.+b acoustoelectric effects 
% 7335 mesoscopic systems 
% 7215R quantum localization 
% 73.20.Fz Weak localization effects (e.g., quantized states) 

Very re\-cent\-ly, sur\-face acous\-tic waves (SAW's) have been used
to study me\-so\-sco\-pic  
systems \cite{Shilton96,Tilke96,Nash96}. 
In ref.~\cite{Shilton96}, the direct acousto-electric current 
induced by a SAW through a single quantum point contact, which can be
considered  
as a quasi-one-dimensional channel of length 0.5~$\mu$m, has been observed. 
In refs.~\cite{Tilke96} and \cite{Nash96}, the SAW attenuation and the 
change of the sound velocity due to arrays of quantum wires and 
quantum dots patterned in a two-dimensional 
electron gas (2DEG) have been measured as a function of a quantizing  
magnetic field. The experiments reported in ref.~\cite{Nash96} include 
dots with a lithographic size as small as 250~nm, their electronic size 
being probably much smaller. 
The measurements of refs.~\cite{Tilke96} and \cite{Nash96} were done on 
isolated dots and wires, {\it i.e.}, no current-carrying 
contacts were attached to the mesoscopic 
systems. This preserves the phase coherence of the electronic wave functions and eliminates 
Coulomb blockade effects. Thus, the SAW method provides a novel tool for non-invasive 
investigations of nanostructures. Experimental studies of isolated mesoscopic systems may also use the 
polarizability of, or the microwave absorption due to small metal 
particles, as recently discussed in refs.\ 
\cite{Efetov96,Noat96} and \cite{Zhou96}, respectively. 
 
In this Letter we calculate the attenuation coefficient, $\Gamma$, 
of SAW's due to an array of quantum dots. We consider dots with an electronic size in 
the nanometer range, {\it e.g.}\ $L=300$~nm, which corresponds to the mean level 
spacing $\Delta \sim 1$~K. Then, at low enough temperatures, the phase breaking rate 
$\tau_\phi^{-1} \ll \Delta$, {\it i.e.}, the energy levels are only slightly broadened 
due to inelastic processes. Since typical SAW frequencies are in the range  
$\omega \sim 10^8 \div 10^9~\text{s}^{-1} = 1 \div 10~\text{mK} \ll \Delta$, the discreteness 
of the spectrum is relevant, requiring the consideration 
of both direct and relaxational absorption processes, to be explained below. 
We present the dependence of $\Gamma$ on the frequency $\omega$, the size $L$ of the 
dots and the temperature $T$, showing that
the relaxational absorption is dominant in a wide range of parameters. 
$\Gamma$ also exhibits weak localization effects, {\it i.e.}, it is sensitive to weak magnetic fields $B$ 
and the spin-orbit scattering rate $\tau_{\text{so}}^{-1}$. This sensitivity 
is dramatically enhanced as compared to a quasicontinuous spectrum ($ \tau_\phi^{-1} > \Delta$ or $\omega \gg \Delta$), 
because the sound absorption resolves 
level correlation effects appearing on a very small energy scale, $\epsilon <\Delta$. 
Direct and relaxational processes involve different time scales of the electronic system, namely 
the phase coherence time $\tau_\phi$ and the energy relaxation time $\tau_\epsilon$, 
suggesting that SAW attenuation measurements may provide information 
on their temperature dependence and magnitude in {\it isolated} mesoscopic systems. 
This could contribute to the clarification of the controversial issue related to the 
significant discrepancies \cite{Bird95,Clarke95,Mittal96} between the theoretical predictions 
for these times and recent experimental results obtained from {\it transport} 
measurements. 
This applies to both the  electron-electron   
interaction and the interaction between electrons and thermal phonons;  
we comment below on   
the relevance of these two interactions for the present considerations. 
 
At sufficiently low temperatures, when the inelastic level broadening 
and the thermal broadening of the distribution function are 
significantly smaller than $\Delta$, the sound absorption depends strongly 
on the existence of narrow pairs of levels whose separations are $\ll \Delta$. 
In general, two absorption mechanisms can be identified \cite{scatt}: 
(i) direct transitions   
between the energy levels involving absorption and emission of   
surface-acoustic phonons, 
and (ii) the absorption due to relaxation processes \cite{Gorter36}.
The latter arise from the periodic motion of the energy levels under the
influence of the external SAW field which, at finite temperatures $T>0$,
leads to a non-equilibrium occupation of the levels.
Energy dissipation  is then due to   
inelastic relaxation mechanisms, which attempt to restore   
instantaneous equilibrium occupancies among the energy levels. Such   
processes were first introduced by Debye in connection with the   
relaxation of a vibrating dipole in a liquid, and then extensively   
discussed in connection with energy dissipation in various physical   
systems. In the following, we call the relaxational absorption simply  
Debye absorption.   
The Debye   
processes involve the occupation relaxation rate of the 
electronic system which is related to the energy relaxation rate $\tau_\epsilon$. 
The direct transitions depend on the phase coherence time $\tau_\phi$.    
In terms of the density matrix formalism, $\tau_\epsilon^{-1}$ and $\tau_\phi^{-1}$
can be identified with the diagonal and the off-diagonal relaxation rates, respectively;
see the discussion in ref.\ \cite{Kamenev95}.
Indeed, the former describes the approach to equilibrium, while the latter is associated
with the destruction of coherence effects, {\it i.e.}, the suppression of the off-diagonal elements
of the density matrix.
 
To describe the energy spectrum of the dots at the scale 
$\epsilon < \Delta$, we use the Random Matrix Theory (RMT) \cite{Mehta67}. 
According to RMT, the statistical properties of the 
spectrum depend only on the global symmetries of the system, such that 
three pure cases can be distinguished: systems with time reversal symmetry (orthogonal ensemble, $B=0$), 
systems with broken time reversal symmetry (unitary ensemble, $B\ne 0$), and systems 
without rotational symmetry (symplectic ensemble, $\tau_{\text{so}}$ small and $B=0$). 
The crossover between these ensembles depends on the energy scale of 
interest, which  may allow to discriminate between different sound attenuation mechanisms, cf.\ below. 
 
In the following, we give a brief derivation of the major results. 
We focus on the case of diffusive dots \cite{ballistic}, where the size of the dots  
exceeds the elastic mean free path, $L>\ell$, and the dimensionless conductance is large, $g\gg 1$. 
For $T<\Delta$, only a narrow pair of levels (say 1 and 2) with   
energetic separation $\epsilon_1-\epsilon_2=\epsilon<\Delta$ 
is of importance for the Debye processes because all other levels, 
up to exponentially small corrections, are completely filled or completely empty;
thus a non-equilibrium occupation does not occur. 
The power absorbed due to these processes by a two-level system   
is \cite{Gorter36}
\begin{equation}\label{qdcetls} 
Q_{\text{D}}(\epsilon)= 
\left(- \frac{\partial f(\epsilon)}{\partial \epsilon} \right) 
\frac{\omega^2\tau_\epsilon(\epsilon)}{1+\omega^2 \tau_\epsilon^2(\epsilon)} 
|M_{11}-M_{22}|^2. 
\end{equation}% 
The matrix elements $M$ are calculated with the eigenfunctions of the states 1 and 2
and the piezoelectric field induced by the SAW. 
(The deformation-potential coupling is negligibly small.) 
The occupation probabilities of the levels 1 and 2 are given by 
$f(\epsilon)=[\exp{(\epsilon/T)}+1]^{-1}$ and $f(-\epsilon)$, respectively. 
Equation  (\ref{qdcetls}) has to be averaged over different realizations of disorder 
in different dots, being associated with statistically independent \cite{tls} variations 
of the eigenfunctions 1 and 2 and the level spacing $\epsilon$. 
We assume that one can average separately over the wave functions in $\tau_\epsilon$ 
and in the difference of matrix elements. The latter quantity is calculated below [eq.~(\ref{amael})] 
and found to be independent of $\epsilon$ in the range of interest. The remaining averaging 
over $\epsilon$ uses the level correlation function 
$R(\epsilon)$ \cite{Mehta67}, that has the limits
\begin{equation} \label{reps} 
R(\epsilon) = \frac{1}{\Delta} \left\{ \begin{array}{lll} 
c_\beta |\epsilon/\Delta |^\beta 
&\text{for} &|\epsilon| \ll \Delta, \\ 
1&\text{for} &|\epsilon| \gg \Delta , \end{array} \right. 
\end{equation}% 
where $c_\beta \simeq 1$, and, as usual, $\beta=1,2$, and 4 for the  
orthogonal, unitary, and symplectic ensembles, respectively. 
As a result (numerical factors of order unity are skipped), 
\begin{equation}\label{qdr} 
Q_{\text{D}}= 
|M_{11}-M_{22}|^2 
\frac{\omega^2\tau_\epsilon(T)}{1+\omega^2 \tau_\epsilon^2(T)} 
R(T) , 
\end{equation}% 
where we have used that the characteristic energy scale in eq.~(\ref{qdcetls})  
is given by $(-\partial f/\partial \epsilon)$ and, hence, is of order $T$. Indeed, 
the energy relaxation rate $\tau_\epsilon^{-1}(\epsilon) \propto \epsilon^p$
depends too 
weakly on $\epsilon$ in order to provide a cut-off on the average over $\epsilon$. 
In particular, one can show \cite{Knabchen97d} that 
inelastic transitions induced by the piezoelectric interaction with thermal phonons 
lead for $\epsilon<T$ to $\tau_\epsilon^{-1}(\epsilon) \propto \epsilon^2$, {\it i.e.}, $p=2$, 
while electron-electron processes are expected to yield $p$ in the range between 1 and 2; see
{\it e.g.} ref.\ \cite{Blanter96c}. 
This is an important difference compared
to the case of 3D metallic particles, where the deformation-potential coupling, 
associated with $p=4$, is dominant (see ref.~\cite{Zhou96}), thus providing a cut-off depending
on $\omega$ and $\tau_\epsilon$. 
 
Let us now turn to the power absorbed due to direct transitions,  
\begin{equation}\label{dre1} 
Q_{\text{dir}} = 
\omega^2 \sum_{m\neq n} \frac{\tau_\phi}{1+\tau^2_\phi(\epsilon_{mn}-\omega)^2} 
\frac{f_n-f_m}{\epsilon_{mn}} |M_{mn}|^2 , 
\end{equation}% 
where $\epsilon_{mn}=\epsilon_m-\epsilon_n$ and $f_n$ measures 
the occupation probability of level $n$.   
Since $Q_{\text{dir}}$ is determined by transitions close to the last occupied level (at $T=0$),
we neglect a possible dependence of $\tau_\phi^{-1}$ on $\epsilon_{mn}$ and
explicitly indicate in the following only its temperature dependence.
Two main contributions to the sum in eq.\ (\ref{dre1}) can be identified: the first is due to
dots where at $T=0$ the last occupied level is separated from the first empty
one by a gap of order $\omega$, while the second arises from dots with gaps of
order $\Delta$.
In the latter case, transitions occur only due to the overlap of the tails of the broadened 
levels. This overlap is strongest for adjacent states, {\it i.e.},  
$\epsilon_{mn}=\epsilon_{m,m\pm 1}\simeq \Delta$. Thus, the level correlation function drops
out and we obtain
\begin{equation}\label{dre2} 
Q_{\text{t}} =  
|M_{12}|^2 (\omega/\Delta)^2 [\tau_\phi(T) \Delta]^{-1} ,
\end{equation} 
where the off-diagonal matrix element $M_{12}$, that is independent of $\epsilon_{12}$, 
is given in eq.\ (\ref{amael}) below.
For dots with gaps as small as $\omega$, 
the level broadening is not relevant and, hence, the Lorentzian 
in eq.~(\ref{dre1}) can be replaced by $\pi\delta(\epsilon_{mn}-\omega)$.
The occupation probabilities of these narrow pairs of levels
are again given by $f(\pm \epsilon)$, cf.\ eq.\ (\ref{qdcetls}). This yields the resonant 
contribution (cf.\ the calculations 
in refs.~\cite{Shklovskii82} and \cite{Sivan86}) 
\begin{equation}\label{dre3} 
Q_{\text{r}}  =  
|M_{12}|^2 \omega \tanh{(\omega /2T)}\, R(\omega) .
\end{equation} 
This result is valid for $\omega, T \ll \Delta $.

The matrix elements in eqs.~(\ref{qdr}), (\ref{dre2}), and (\ref{dre3})   
are given by  
$M_{mn} = \left\langle m| 
V_\omega \exp{(i\vec{q}\vec{r})} | n\right\rangle $, 
where $V_\omega=\gamma_q/({\cal L}\varepsilon)$ represents the screened piezoelectric potential 
arising from the SAW. The interaction vertex $\gamma_q$ was calculated in ref.~\cite{Knabchen96}, 
and ${\cal L}^2$ is the normalization area in the plane of the quantum dots. 
The screening of the SAW potential is described by the dielectric function $\varepsilon$
that is very much dependent on the fact that we are studying a 2D system:
First, the piezoelectric field penetrates the quantum dot completely and, hence, 
$\varepsilon$ may be taken to be $\vec{r}$-independent. The sound attenuation thus clearly probes
the behaviour of the electron wavefunctions within the whole dot.
In 3D metallic particles only the electrons in a distance up to the screening length from the
surface are subject to an external field \cite{Zhou96}, since the field decreases exponentially
towards the interior of the particle.
Secondly, the Coulomb interaction between different dots (which is in leading
approximation a dipole-dipole interaction decaying like $r^{-3}$)
is not very effective for a 2D array of dots. Indeed, experiments have shown \cite{Dahl93}
that the interdot coupling leads even in dense arrays to a change of the local electric field
of no more than 25\%. This is again very different from the case of a 3D system containing metallic grains, where
their mutual coupling has to be taken into account with care; cf.\ the discussion
in ref.\ \cite{Sivan86}. Neglecting the interdot coupling,
the calculation of $\varepsilon$ coincides in the linear screening regime
with the evaluation of the density response function. Interestingly, while
the SAW attenuation probes the properties of a few levels close
to the last occupied level at $T=0$, the density response function is
determined by (virtual) transition processes with large
energy transfers $|\epsilon_{mn}| \sim E_c=D/L^2=g\Delta \gg \Delta$,
where $D$ is the diffusion coefficient of the electron gas. Consequently,
the screening is neither sensitive to the symmetry of the system ($\beta=1,2,4$)
nor depends on the discreteness of the spectrum. It is therefore natural that the
dielectric function reduces to its familiar expression for a diffusive
system, taken in the limits $\omega \rightarrow 0$ and $q \sim L^{-1}$, {\it i.e.},
$\varepsilon \simeq L/ a_B\gg 1$ \cite{Knabchen97a},
where $a_B$ is the Bohr radius.

For typical sound frequencies, $qL \ll 1$, and, hence, the matrix 
element simplifies to   
$M_{mn} = i V_\omega \vec{q} \left\langle m|\vec{r}| 
n\right\rangle$. [For $m=n$  
the first expansion term is canceled after taking the difference of diagonal matrix 
elements, cf.\ with eq.~(\ref{qdr}), {\it i.e.}, we have again to consider the matrix element of $\vec{r}$.]  
Disorder averaged products of these matrix elements 
can be calculated using, {\it e.g.}, quasi-classical methods \cite{Gorkov65,Sivan86} or wavefunction 
correlators derived within the supersymmetry approach \cite{Blanter96}. 
We obtain 
\begin{equation}\label{amael} 
|M_{11}-M_{22}|^2 \simeq |M_{12}|^2 \simeq (\gamma_q  q a_B)^2/({\cal 
L}^2 g). 
\end{equation}% 
This result is valid for energy differences $|\epsilon_1-\epsilon_2| \le E_c$. 
 
The SAW attenuation coefficient $\Gamma$ is related to the energy loss $Q$  
according to
$ \Gamma= {\cal N}{\cal L}^2 Q /(\hbar \omega s)  $, where ${\cal N}$ is the areal density
of quantum dots and $s$ the sound velocity. 
Collecting the above results for $Q$ and 
the matrix elements yields 
\begin{eqnarray} 
\Gamma_{\text{D}} & =&  
A\,
\frac{\omega}{s} 
\frac{\omega \tau_\epsilon(T)}{1+\omega^2\tau^2_\epsilon(T)}  
\hbar \omega R(T) , \label{gamd} \\ 
\Gamma_{\text{t}} & =&  
A \,
\frac{\omega}{s}
\left( \frac{\hbar\omega}{\Delta}\right)^3 \frac{1}{\omega \tau_\phi(T)} , 
\label{gamt} \\ 
\Gamma_{\text{r}}& =& 
A\,
\frac{\omega}{s} 
\tanh{(\frac{\hbar\omega}{2T})}  
\hbar \omega R(\hbar\omega) , \label{gamr} 
\end{eqnarray} 
for the Debye attenuation, the attenuation due to the tails of the broadened levels, 
and the resonant attenuation, 
respectively, and $A=({\cal N}/g) (a_B \gamma_q/\hbar s)^2 $. 
Let us now substitute some realistic parameters 
in order to 
estimate the relative magnitude of these terms. In GaAs/Al${}_x$Ga${}_{1-x}$As heterostructures,
$\gamma_q/s \approx 1$ \cite{Knabchen96}. 
For the parameters $L=300$~nm and $\Delta=1$~K introduced above, a value of 
$g\sim 5$ results in $D \sim 10^{-2}$~m${}^2$s${}^{-1}$. We put $T\sim 0.1$~K, and 
assume $\tau_\phi^{-1}(T) \sim 10^9$~s${}^{-1}$ and $\tau^{-1}_\epsilon(T)\sim 10^8$~s${}^{-1}$. 
The phase breaking rate is about one order of magnitude smaller than the one recently derived 
from {\it transport} experiments \cite{Bird95,Clarke95} with ballistic quantum dots of size $L\sim 2~\mu$m. 
This reduction of the dephasing seems to be reasonable since we are considering isolated dots without leads.
The energy relaxation rate is only sensitive to inelastic scattering events with a substantial 
energy transfer and, hence, should be always smaller than $\tau_\phi^{-1}$ \cite{Altshuler82a}. 
Then, for typical SAW frequencies, we find  
the inequalities $\Delta > T > \tau_\phi^{-1} 
\ge \omega  \ge \tau^{-1}_\epsilon$, where 
$T> \omega$ is most probably always fulfilled. 
Based on these parameters we obtain $\Gamma_{\text{D}} \gg \Gamma_{\text{r}} \ge \Gamma_{\text{t}}$ for $\beta=1$, and 
$\Gamma_{\text{D}} \gg \Gamma_{\text{t}} \gg \Gamma_{\text{r}}$ for $\beta=2$. The Debye absorption remains dominant if 
$\omega\tau_\epsilon(T)$ varies in the range $10^{-3}-10^2$. Thus, the measurement of 
$\Gamma(T)=\Gamma_{\text{D}}(T)$ [eq.~(\ref{gamd})] at a fixed SAW frequency accesses the temperature 
dependence of the energy relaxation time $\tau_\epsilon(T)$ in systems with or without
time reversal symmetry. 
Very interesting is the case of broken rotational symmetry, $\beta=4$, because $\Gamma_{\text{D}}$ may now decrease 
below $\Gamma_{\text{t}}$ [$\gg \Gamma_{\text{r}}$], permitting the measurement of both $\tau_\epsilon(T)$ 
and $\tau_\phi(T)$ at different frequencies. 
Generally, $\Gamma_{\text{D}}$ is [depending on the density of the dot lattice and other parameters] 
about 2-3 orders of magnitude smaller than the maximum SAW attenuation 
due to an extended 2DEG, which should be within the experimental 
resolution.

This discussion has shown that the global symmetries of the system (described by the
parameter $\beta$) determine directly which of the various contributions
to the attenuation coefficient [eqs.\ (\ref{gamd})--(\ref{gamr})] is dominant.
The parameter $\beta$ enters via the level correlation function $R(\epsilon)$, eq.\ (\ref{reps}),
if energy scales smaller than $\Delta$ are relevant. This is not the case for
the tail absorption so that $\Gamma_{\text{t}}$ is essentially independent of the symmetries.
The crossover from the time reversal invariant case ($\beta=1$) to that with broken time reversal symmetry
($\beta=2$) is achieved by applying a magnetic field $B\gg B^*$. The very small threshold field
is given by $B^*\simeq (\Phi_\circ/L^2) \sqrt{\epsilon^*/E_c}$, where
$\Phi_\circ$ is the flux quantum and $\epsilon^*$ is the typical energy scale of the transitions.
Since $\Gamma_{\text{D}}$ and
$\Gamma_{\text{r}}$ are associated with $\epsilon^*=T$ and $\epsilon^*=\omega$, respectively,
the threshold fields are different,
\begin{equation}\label{flux}
B^*_{\text{ D}} / B^*_{\text{r}} = \sqrt{T/\omega} .
\end{equation}%
Similarly, the rotational symmetry can be broken by increasing the spin-orbit scattering
rate $\tau^{-1}_{\text{so}}$ above $\epsilon^*$. One should note that the variation of $\beta$ results
in a significant change of the magnitude of $\Gamma$. In this sense, the dependence of the sound absorption
on weak magnetic fields and spin-orbit scattering is much more pronounced in the discrete
spectrum limit than in the quasicontinuous case. In the latter, this dependence arises solely
from weak localization {\it corrections} to the conductance $g$ (entering $A$) \cite{Knabchen97a}.
In the present case, these corrections can safely be ignored as compared to the effects
associated with $R(\epsilon)$.

In summary, we have considered the attenuation of surface-acoustic waves due to an array of 
quantum dots in the discrete spectrum limit, arising from  both direct and Debye processes. 
It has been shown that non-invasive measurements of $\Gamma$ can yield the temperature 
dependence of both the energy and the phase breaking rate of an isolated electronic system. 
The sensitivity to weak magnetic fields or the spin-orbit scattering (``weak localization effects'') 
is greatly enhanced as compared to the case of a continuous spectrum. 
It might be possible to observe the transition from the thermal-phonon to the electron-electron dephasing,  
which is known to occur at around 1~K \cite{Mittal96}. 
We hope that these calculations will stimulate further experimental 
investigations in this field. 
 
\stars  
 
Financial support by the German-Israeli Foundation, the 
Fund for Basic Research administered by the Israel Academy 
of Sciences and Humanities, and the Deutsche Forschungsgemeinschaft (A.~K.) 
is gratefully acknowledged. 
We thank Y.\ Gefen, Y.\ Imry, and A.\ Wixforth for valuable discussions.

\newcommand{\noopsort}[1]{} \newcommand{\printfirst}[2]{#1} 
  \newcommand{\singleletter}[1]{#1} \newcommand{\switchargs}[2]{#2#1}


\begin{thebibliography}{10}
\bibitem{scatt} 
In principle, the {\it elastic} scattering of surface phonons may show up in 
the experiments as an attenuation mechanism ($\Gamma_{\text{sc}}$) that is not affected by 
the fact that $\omega \ll \Delta$. 
We find $\Gamma_{\text{sc}} \propto \omega^5$ which is small unless $\omega$ becomes very large.
\bibitem{ballistic} 
We believe that our results remain valid (by order of magnitude) in the ballistic regime 
if $E_c=D/L^2$ is replaced by the inverse time of flight $v_F/L$.
\bibitem{tls} 
This is a general statement of RMT. Note, however, 
that this decoupling is not valid for a localized tunneling two-level system, where the overlap between 
the states 1 and 2 determines both their splitting and the transition matrix elements.
\end{thebibliography}
\end{document}